\documentclass[referee]{aa} 

\usepackage{graphicx}
\usepackage{txfonts}

\begin{document}

\title{Cold prominence materials detected within magnetic clouds during 1998-2007}
\author{Jiemin Wang, Hengqiang Feng, Guoqing Zhao}
\offprints{Hengqiang Feng}
\institute{Institute of space physics, Luoyang Normal University, Luoyang 471934, China\\
              \email{fenghq9921@163.com}
            }

\abstract
{Coronal mass ejections (CMEs) are intense solar explosive eruptions, and they are frequently correlated with prominence eruptions. Previous observations show that about $70\%$ of CMEs are associated with prominence eruptions. However, there are only a handful of reported observations of prominence plasma materials within interplanetary CMEs (ICMEs), which are the interplanetary manifestations of CMEs. Moreover, approximately $4\%$ of ICMEs exhibit the presence of prominence materials, and approximately $12\%$ of magnetic clouds (MCs) contain prominence materials. }
{We aim to comprehensively search for cold prominence materials in MCs observed by the Advanced Composition Explorer (ACE) spacecraft during 1998-2007.}
{Using the criteria of unusual $O^{5+}$ and (or) $Fe^{6+}$ abundances, we examined 76 MCs observed by ACE during 1998-2007 to search for cold prominence materials.}
{Our results revealed that out of the 76 MCs, 27 ($36\%$) events contained prominence material regions with low-charge-state signatures. }
{Although the fraction is still lower than the approximately $70\%$ of CMEs associated with prominence eruptions, it is much higher than $12\%$. The unusual $O^{5+}$ and (or) $Fe^{6+}$ abundances may be simple and reliable criteria to investigate prominence materials in the interplanetary medium.}


\authorrunning{Wang, Feng \& Zhao}
\titlerunning{Cold materials detected within magnetic clouds}

\maketitle

\section{Introduction}

Coronal mass ejections (CMEs) are an important kind of solar eruption activity. Large amounts of plasma and magnetic fields are ejected from the corona in the form of a CME. CMEs often show different morphologies in coronagraph images, such as loop, streamer separation, and filled bottle  (Munro \& Sime 1985). Many CMEs are observed by coronagraphs as classical three-part structures: a bright leading loop, a dark cavity, and a bright dense core in the cavity (Kahler 1987). CMEs were found to be associated with two other solar activities: solar flares and prominence eruptions. About 40 years ago, Munro et al. (1979) reported that more than $70\%$ of CMEs are associated with prominence eruptions, $50\%$ with prominence eruptions solely (without flare), and $40\%$ with flares. On the other hand, $72\%$ of solar prominence events are also clearly associated with CMEs (Gopalswamy et al. 2003).

\begin{figure*}
\sidecaption
\resizebox{\hsize}{18cm}{\includegraphics[width=10cm]{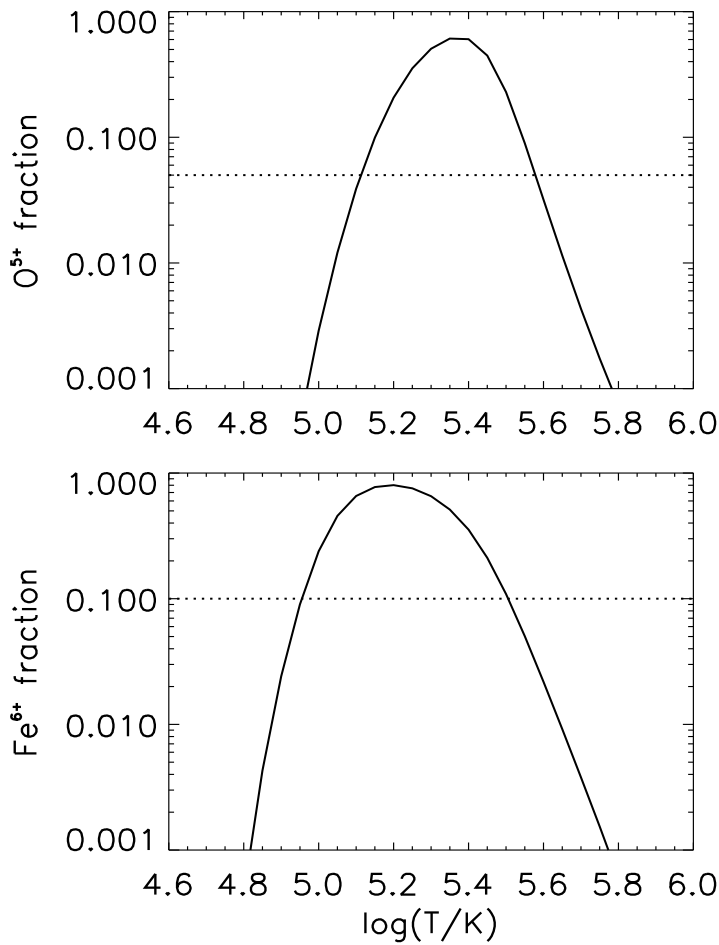}}
\caption{Top panel: The $O^{5+}$ fraction of total oxygen ions versus solar atmospheric temperatures (T). The horizontal dotted line with a value of 0.050 denotes the an  abundance criterion for unusual abundance in the present paper. Bottom panel: The $Fe^{6+}$ fraction of the total oxygen ions versus the solar atmospheric temperatures. The horizontal dotted line with a value of 0.100 denotes the an abundance criterion for unusual abundance, where K indicates the units of Kelvin. The figure is reproduced based on the database and IDL programs provided by the CHIANTI web site.}
\end{figure*}

As mentioned above, prominence eruptions are most frequently associated with CMEs (e.g., Webb \& Hundhausen 1987). However, there are only a handful of reported observations of prominence plasma materials within interplanetary CMEs (ICMEs), which are interplanetary manifestations of CMEs. Burlaga et al. (1998) reported a halo CME observed on January 6, 1997 by the Large Angle Spectrometric Coronagraph (LASCO) on board the Solar and Heliospheric Observatory (SOHO); halo CMEs are CMEs originating from the solar disk and quickly expanding to a projected size larger than the occulting disk of coronagraphs (Howard et al.1982; Thompson et al. 1998). Moreover, the halo CME is associated with a disappearing filament. The corresponding related ICME was a magnetic cloud (MC), which was introduced by Burlaga et al. (1981, 1990) to empirically define a special subset of ICMEs using three necessary characteristics: smooth rotation of magnetic field vector, enhanced magnetic field strength, and low proton temperature. The remnant prominence materials were first identified within the MC through high-density and enhanced helium abundance origin with low proton temperatures and low ionic charge states by Burlaga et al. (1998). To date, the reported ICMEs embedded prominence plasma materials are still rare (Gopalswamy et al. 1998; Skoug et al. 1999; Lepri \& Zurbuchen 2010; Sharma \& Srivastava 2012; Sharma et al. 2013). Lepri \& Zurbuchen (2010) systematically investigated solar wind data observed by the Advanced Composition Explorer (ACE) during 1998-2007 to search for cold prominence material in ICMEs. They found that only about $4\%$ of 283 ICMEs exhibited signatures of prominence plasma, and approximately $12\%$ of MCs contained prominence materials. The fraction is much less than that of CMEs, of which $70\%$ were clearly associated with prominence eruptions. They also gave two possible explanations for the great difference. One is that their selection criteria were highly restrictive; many cold prominence material regions in ICMEs may have been missed. The other possibility is that the low charge states of prominences were heated by accompanied flares near the Sun, and the low-charge-state signatures in the prominence plasmas have been erased when the ICMEs reached about 1 AU.

In this study, we will comprehensively search for cold prominence materials in MCs observed by the ACE spacecraft during 1998-2007, using a broader threshold selection criteria to identify remnant prominence materials.
Our results revealed that out of the 76 MCs, 27 ($36\%$) events contained prominence material regions with low-charge-state signatures. Although the fraction is still lower than that of CMEs, of which approximately $70\%$ are associated with prominence eruptions, it is much higher than $12\%$.

\section{ Methodology and data}

The relatively low temperature and high density are two essential observational characteristics of solar prominences. Therefore, the remnant prominence materials within MCs have often been identified using their low temperatures, low ionic charge states, and high-proton density signatures (e.g., Burlaga et al. 1998; Sharma \& Srivastava 2012; Sharma et al. 2013). However, ICMEs (including MCs) usually have outstanding characteristics, such as  enhanced magnetic fields. Thus, total pressures within ICMEs may be higher than the outside, and the ICMEs will expand quickly when they depart from the Sun. If ICMEs have local dropped magnetic field regions, then their expanding speed will be slower than that of the adjacent parts. Consequently, the dropped magnetic field regions may become a relatively high-proton density region. In the same way, the original high-density regions can appear, and new low-temperature regions can also be formed at 1 AU through non-uniform expansion of ICMEs. Therefore, the high density and low temperature signatures of ICMEs may not be reliable evidence for prominence materials. On the other hand, the ion charge states are frozen-in near the Sun when ionization and recombination timescales are larger than the solar wind ion expansion time, since the coronal electron density decreases with increasing distance from the Sun (Heidrich-Meisner et al. 2016). In steady state solar atmosphere conditions, charge states of $Fe$ and $O$ freeze-in within four solar radii (Hundhausen et al. 1968a, 1968b; Bame et al. 1974; Buergi \& Geiss1986). For these reasons, the ion charge states in ICMEs can still reflect their temperature history (Richardson \& Cane 2004; Heidrich-Meisner et al. 2016; Wang \& Feng 2016). Therefore, the low ionic charge states may be a reliable signature to identify prominence materials in the interplanetary medium. The previous reported cold prominence materials within ICMEs have exhibited low charge states. For example, both Burlaga et al. (1998) and Skoug et al. (1999) used the presence of $He^{+}$ to identify cold prominence materials within MCs. The prominence material region of Burlaga et al. (1998) also contained the unusual charge states of $O^{5+}$ and $Fe^{5+}$. Lepri \& Zurbuchen (2010) focused on the low charge states of $Fe^{4+}$, $C^{2+}$, and $O^{2+}$. Yao et al. (2010) also found that the thermal ion velocity distribution functions are more isotropic and appear to be colder than in the other regions of the MCs. In the present study, we will use the unusual charge states of $O^{5+}$ and (or) $Fe^{6+}$ to identify the remnant prominence materials. In the million-degree typical coronal atmosphere, $O^{5+}$ abundance (the $O^{5+}$ fraction of the total oxygen ions) and $Fe^{6+}$ abundance (the fraction of the total iron ions) should be very small. Thus, unusual $O^{5+}$ and (or) $Fe^{6+}$ abundances may be taken as indicators to identify prominence materials in the interplanetary medium. Dere et al. (1997) and Landi et al. (2013) established and improved the CHIANTI database, which can calculate the charge state distributions of the elements such as iron and oxygen in the solar atmosphere. All data and programs are also freely available at the CHIANTI web site, http://www.chiantidatabase.org. Figure 1 shows the changes of the $O^{5+}$ fraction of the total oxygen ions versus the solar atmospheric temperatures (top panel), and the $Fe^{6+}$ fraction of the total iron ions versus the freezing-in temperatures (bottom panel). Figure 1 was reproduced based on the database and interactive data language (IDL) programs provided by the CHIANTI web site. These fractions have been calculated under equilibrium conditions in the solar atmosphere, they may not be entirely suitable for the CME or prominence plasma. However, the frozen-in ion fractions information within ICMEs still can reflect their historical temperature near the Sun (Gilbert et al. 2012). According to the curve of the $O^{5+}$ fraction in Figure 1, the corresponding source freezing-in temperature of higher $O^{5+}$ fractions is much lower than typical coronal temperatures. Therefore, the higher $O^{5+}$ fractions ($\geq0.05$) should not be generally present in the usual interplanetary solar wind. If the $O^{5+}$ fractions in the interplanetary medium are more than or equal to 0.05, then we define such states as unusual $O^{5+}$ abundance. The freezing-in temperature range of the unusual $O^{5+}$ abundance is approximately $1.3-3.8\times10^{5} K$. We also define unusual $Fe^{6+}$ abundance as $Fe^{6+}$ fractions more than or equal to 0.10, and the corresponding freezing-in temperature in the solar atmosphere is approximately $0.9-3.2\times10^{5} K$. Hence, the freezing-in temperatures of unusual $O^{5+}$ and $Fe^{6+}$ abundances are much lower than the usual coronal temperatures but are consistent with expected temperatures of erupting prominences (Burlaga et al. 1998; Heinzel et al. 2016). Therefore, the unusual $O^{5+}$ and (or) $Fe^{6+}$ abundances may be appropriate indicators to identify prominence materials within MCs.
\begin{figure*}
\sidecaption
\resizebox{\hsize}{12cm}{\includegraphics[width=12cm]{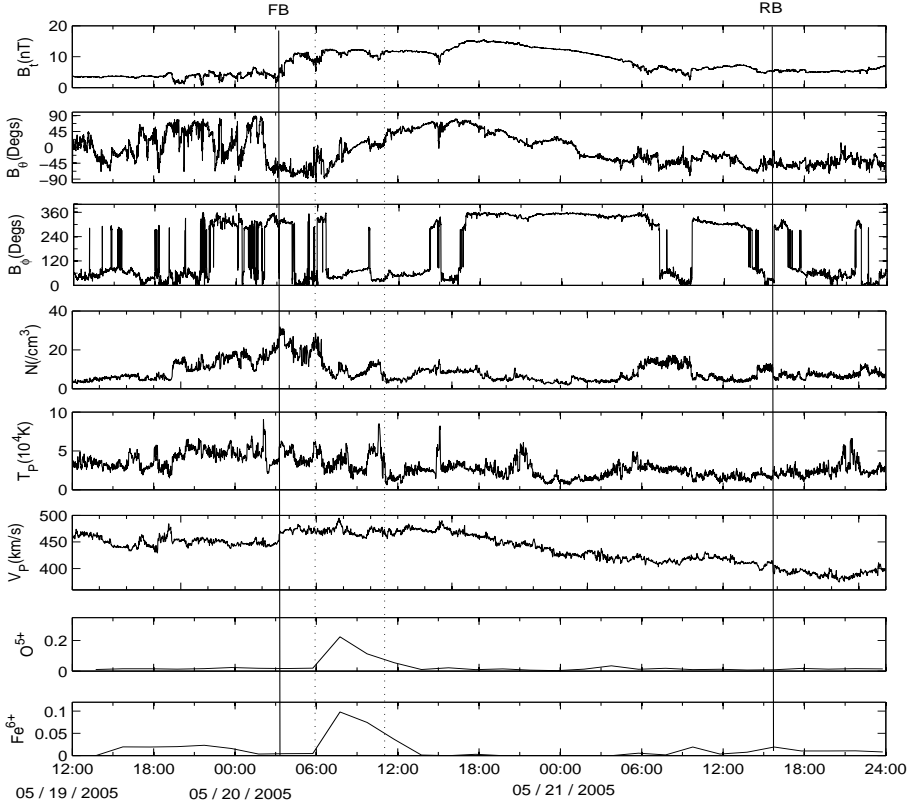}}
\caption{Magnetic field and plasma data: the $O^{5+}$ and $Fe^{6+}$ abundances during the MC on May 19-21, 2005 passage. The two vertical solid lines denoted by FB and RB indicate the front and rear boundaries of the MC, respectively. The cold plasma region was labeled by two vertical dot lines.}
\end{figure*}

To assess the above thresholds of the unusual $O^{5+}$ and $Fe^{6+}$ abundances, we also examined the $O^{5+}$ and $Fe^{6+}$ abundances of the solar wind in the month of June in every year from 1998 to 2007;  the month was chosen randomly in the first year but it was determinate in the following years. As the average timescale of the 76 MCs is approximately 38 h, each June from 1998 to 2007 was then divided into many time periods, and each time periods lasts for 38 hours. During this process, the time periods with ICME (MC) occurring were excluded. Finally 159 time periods were analyzed, and results showed that unusual $O^{5+}$ or $Fe^{6+}$ abundances were observed only in three time periods of the usual interplanetary solar wind. The three time periods are as follows: (1) interval from 22:00 UT on June 8 to 12:00 UT on June 10, 1999, with maximum $O^{5+}$ fraction value of 0.059; (2) interval from 12:00 UT on June 10 to 02:00 UT on June 12, 1999, with maximum $Fe^{6+}$ fraction value of 0.110; and (3) from 02:00 UT on June 12 to 16:00 UT on June 13, 2007, with maximum $O^{5+}$ fraction value of 0.057. However, we examined 76 MCs observed by ACE during 1998-2007 and found that 27 events contained the \textsc{unusual $O^{5+}$ and/or $Fe^{6+}$ abundances}. This indicates that the unusual $O^{5+}$ and/or $Fe^{6+}$ abundances are usually related to MCs and presented in identified filament materials (Burlaga et al. 1998; Gopalswamy et al. 1998). Therefore, the unusual $O^{5+}$ and/or $Fe^{6+}$ abundances may be simple and reliable criteria to identify prominence materials in the interplanetary medium.

As mentioned above, Lepri \& Zurbuchen (2010) searched for cold prominence materials in ICMEs during 1998-2007 by using the low charge states of $Fe^{4+}$, $C^{2+}$, and $O^{2+}$. They found that only approximately $12\%$ of MCs exhibited prominence plasma. As Lepri \& Zurbuchen (2010) pointed out, their selection criteria were highly restrictive, and the freezing-in temperatures of these charge states were low: $0.6-2.0\times10^{5} K$. Thus, many cold filament material regions in MCs (ICMEs) may have been missed. In comparison with the selection criteria of Lepri \& Zurbuchen (2010), the freezing-in temperature ranges of unusual $O^{5+}$ and $Fe^{6+}$ are about $1.3-3.8\times10^{5} K$ or $0.9-3.2\times10^{5} K;$ both the lower limit temperature and higher limit temperature are greater than that ($0.6-2.0\times10^{5} K$) of Lepri \& Zurbuchen (2010). That is to say, if the freezing-in temperature of prominence materials within ICMEs is too low, the filament material may be missed by our criteria; if the freezing-in temperature of prominence materials within ICMEs is relative high, the filament material may be missed by the criteria of Lepri \& Zurbuchen (2010). We examined all the 11 cold material regions reported by Lepri \& Zurbuchen (2010), and all these cold prominence materials can be identified using our criteria (unusual $O^{5+}$ and (or) $Fe^{6+}$ abundances). It may indicate that the freezing-in temperature of prominence materials within ICMEs is not too low to miss our criteria. Therefore, to reduce the chances of cold prominence materials within MCs being missed, we will search for cold plasma by using the unusual $O^{5+}$ and (or) $Fe^{6+}$ abundances.

In this study, the $O^{5+}$ and $Fe^{6+}$ fraction data were obtained from the Solar Wind Ion Composition Spectrometer (SWICS) on the ACE spacecraft, and their time resolution is 2 h. The instruments and data analysis procedure were described by Gloeckler et al. (1998). We also examined the magnetic field and plasma data during 1998-2007, and identified 76 MCs in total. These MCs were identified with the three necessary characteristics mentioned in Section 1. In addition, we also referred to the ICME lists published by Richardson and Cane (http://www.srl.caltech.edu/ACE/ASC/DATA/level3/icmetable2.htm). We examined all 76 MCs to search for prominence materials.

\section{Observations}

To search for cold prominence material, we examined $O^{5+}$ and $Fe^{6+}$ fractions of the 76 MCs during 1998-2007, and found that 27 of 76 MC events were embedded prominence materials. In this section, we introduce two MCs as examples to exhibit their cold material regions. One is the MC in May 19-21, 2005, in which we found cold materials near the front boundary. Lepri \& Zurbuchen (2010) also found prominence imprints with low charge states in the event. The other is the MC in March 4-6, 1998, in which we found cold materials near the rear boundary.

\begin{figure*}
\sidecaption
\resizebox{\hsize}{12cm}{\includegraphics[width=12cm]{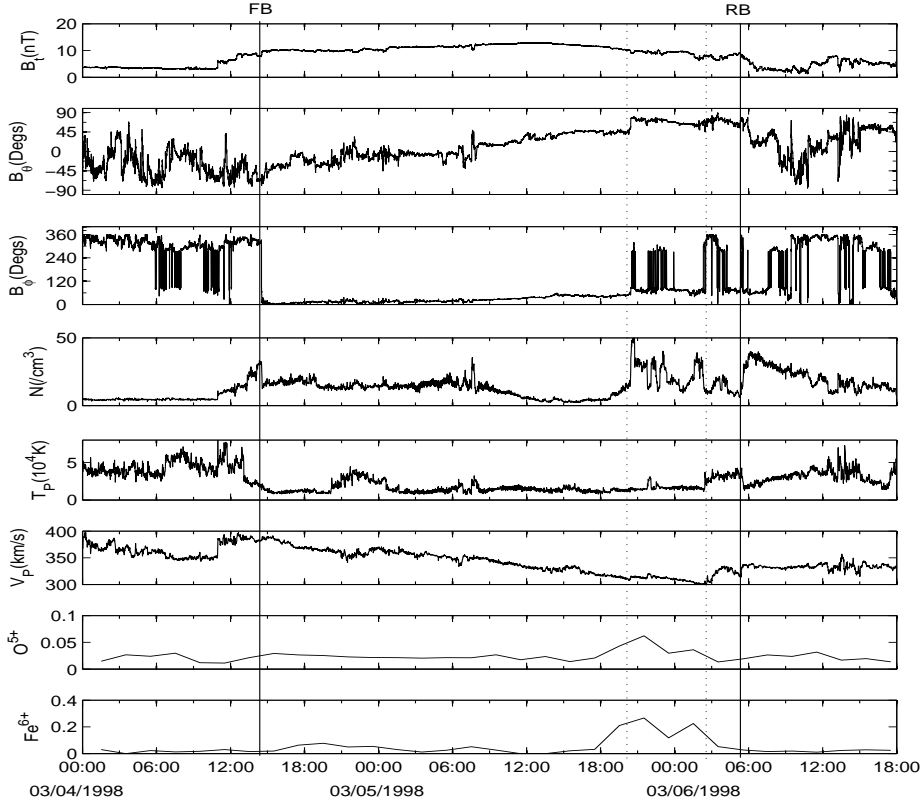}}
\caption{Magnetic field and plasma data: the $O^{5+}$ and $Fe^{6+}$ abundances during the MC on March 4-6, 1998 passage. The meanings of vertical solid and dotted lines are the same as in Figure 2.}
\end{figure*}

Figure 2 shows the magnetic field data in the Geocentric Solar Ecliptic (GSE) angular coordinates, the proton density, proton temperature, proton speed, and $O^{5+}$ and $Fe^{6+}$ fractions during the May 19-21, 2005 MC passage. The event has the classic MC signatures: the total magnetic fields are dramatically higher than the background solar wind, the latitude angle $\theta$ rotates smoothly, and the proton temperatures are low. In Figure 2, the $O^{5+}$ fraction curve clearly has an apparent enhanced region at 05:44-13:45 UT on May 20, and the $O^{5+}$ fractions increased from 0.018 to 0.224. The $Fe^{6+}$ fraction curves also display an enhanced region in the corresponding part, and its maximum value reaches 0.098. According to the above mentioned criteria, this cold plasma region, which is labeled by two vertical dot lines, is identified as prominence materials. In addition, based on the $O^{5+}$ fraction data shown in Figure 1, the estimated freezing-in temperature of 0.224 is $1.66\times10^{5} K$ or $3.39\times10^{5} K$. The estimated freezing-in temperature of the maximum value (0.098) of the $Fe^{6+}$ fraction is $0.93\times10^{5} K$ or $3.24\times10^{5} K$. Therefore, the temperatures of erupting prominences may be approximately $3.24-3.39\times10^{5} K$. Lepri \& Zurbuchen (2010) also identified the cold plasma region as embedded prominence plasma materials. From Figure 2, one can find that the proton density curve has a slight enhancement in the prominence material region. However, the proton temperature curve also has a slight enhancement in the same location. In addition, both the lowest proton temperature location and the highest proton density location are out of the prominence material region. These observations also indicate that the high-density and low-temperature signatures are not reliable evidence for the identification of prominence materials.

\begin{figure*}
\sidecaption
\resizebox{\hsize}{12cm}{\includegraphics[width=12cm]{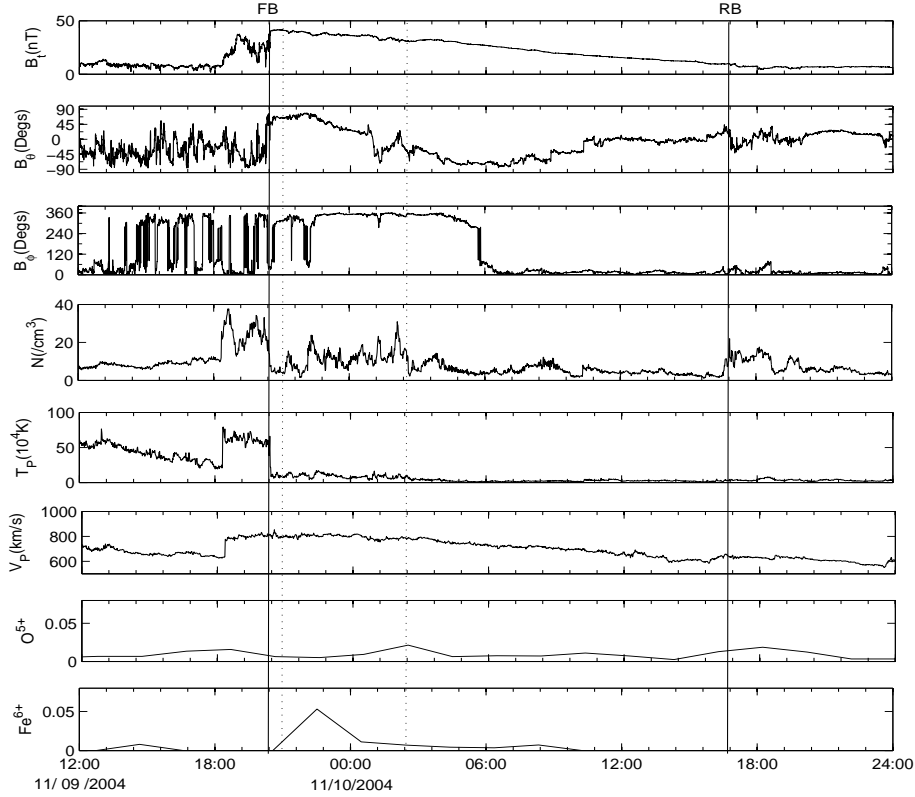}}
\caption{Magnetic field and plasma data: the $O^{5+}$ and $Fe^{6+}$ abundances during the MC on November 9-10, 2004 passage. The meanings of vertical solid and dotted lines are the same as in Figure 2.}
\end{figure*}

Figure 3 shows the MC during March 4-6, 1998. The event exhibits all three typical signatures: smooth rotation of the magnetic field vector, enhanced magnetic field strength, and low proton temperature. The velocity curve shows that the MC has a decreasing velocity profile, indicating that it is still expanding. From Figure 3, we can find that the $O^{5+}$ and $Fe^{6+}$ abundance curves have apparent enhancements before the rear boundary of the MC. Within the enhancement region, the maximum value of the $O^{5+}$ abundance is 0.062, and based on data displayed in Figure 1, its freezing-in temperature is approximately $1.35\times10^{5} K$ or $3.38\times10^{5} K$.  From the data in Figure 1, the maximum value of the $Fe^{6+}$ abundance is 0.266, and its freezing-in temperature is approximately $1.03\times10^{5} K$ or $2.74\times10^{5} K$. These possible freezing-in temperatures of the $O^{5+}$ and $Fe^{6+}$ abundances are consistent with those of erupting prominences. In addition, the proton density curve clearly shows an apparent enhanced region between 20:14 UT on March 5 and 02:30 UT on March 6, 1998, and the proton density increased from 5 $cm^{-3}$ to 60 $cm^{-3}$. From the magnetic component curves, all three magnetic components jump around the high-density region, which may imply that this region is a distinct part. In summary, this high-density region might be a prominence material.

\section{Results and discussion}

According to the criteria defined in Section 2, we examined the 76 MCs during 1998-2007 and found that 27 MCs contained 29 cold material regions, which are presented in Table 1. The second and third columns show the front and rear boundaries of MCs, respectively. The fourth and fifth columns present the start and end times of the cold material regions. The sixth and seventh columns show the maximum values of $O^{5+}$ and $Fe^{6+}$ fractions within the cold material regions. From Table 1, we can find that 13 cold material regions not only have an unusual $O^{5+}$ abundance but also an unusual $Fe^{6+}$ abundance, 12 cold material regions only have the unusual $O^{5+}$ abundance, and four cold material regions only have the unusual $Fe^{6+}$ abundance. One can note that the estimated freezing-in temperature range of the unusual $O^{5+}$ and $Fe^{6+}$ abundances is approximately consistent, but more than half of cold material regions have the unusual $O^{5+}$ or $Fe^{6+}$ abundances. This could be due to the complex freezing-in process of ion charge states near the Sun. The ionization state is not only temperature-dependent. The various ion species freeze-in at different altitudes, and deviations are expected (Bame et al. 1974). It is also possible that the $O^{5+}$ ions had frozen-in at a low altitude; most $Fe^{6+}$ ions were heated and ionized as they moved through the corona, and then the $Fe^{6+}$ ions froze-in at a high altitude. However, either of these two unusual ion abundances can still reflect their low temperature history in the solar source regions. We also examined the spatial distribution of cold material within the MCs. The results show that in 11 ($38\%$) of the events, cold materials were detected in the front part of the MCs; in six ($21\%$) of the events, cold materials appeared in the middle part of the MCs; and in 12 ($41\%$) of the events, cold materials were observed in the rear part of the MCs. The fractions were very consistent with the spatial distribution of cold materials within the ICMEs provided by Lepri \& Zurbuchen (2010). The cold prominence materials may exist in the rear part of CMEs when they depart from the Sun. As CMEs propagate from the Sun to the interplanetary medium, the cold prominence materials can move to the front parts due to strong azimuthal flows within ICMEs (MCs) (Manchester et al. 2014).

In Section 1, it was mentioned that Lepri \& Zurbuchen (2010) had searched cold materials in the ICMEs by using strict selection criteria and found that only 11 ICMEs exhibit cold materials. Among the 11 ICMEs, nine cases were MCs. That is to say, approximately $4\%$ of ICMEs exhibited the presence of prominence materials, and approximately $12\%$ of MCs contained prominence materials. As mentioned above, we identified prominence plasma by using related less-restrictive criteria. Cold prominence materials were detected in 27 ($36\%$) of the 76 MCs. Although the fraction was much higher than $12\%$, it was still lower than that of CMEs, of which approximately $70\%$ are associated with prominence eruptions. The difference may be caused by four possible factors. The first possible factor is that we identified prominence plasma by related less-restrictive criteria, but some cold prominence materials in MCs may still have been missed. For example, Figure 4 shows a MC during November 9 and 10, 2004. The $Fe^{6+}$ abundance curve had an apparent protuberance after the front boundary of the MC, and its maximum value reached 0.073, which indicates that its source region should be more than $3.5\times10^{5} K$ or less than $0.9\times10^{5} K$. The $O^{5+}$ abundance kept a low state during this period, and its maximum value was only 0.021. Its source region should be more than $4.3\times10^{5} K$ or less than $1.2\times10^{5} K$. The possible source temperatures of the high-density region were still consistent with that of erupting prominences. Around the relatively high $Fe^{6+}$ abundance region, proton density was also increased evidently. Therefore, the relatively high $Fe^{6+}$ abundance region might also be a filament material. As Lepri \& Zurbuchen (2010) pointed out, the second possible factor is that the low charge states of prominences were heated by accompanied flares, and the low-charge-state signatures in the prominence plasmas have been erased when the CMEs reached to about 1 AU. The third possible factor is that a CME may be associated with a prominence eruption, but the prominence materials are not embedded in the CME. In fact, many prominences develop a sideways rolling motion (Martin 2003; Gopalswamy et al. 2000, 2015; Gopalswamy \& Thompson 2000; Simnett 2000; Panasenco \& Martin 2008; Panasenco et al. 2011, 2013). Panasenco et al. (2011) reported a CME on December 12, 2008, which traveled in a trajectory different from its accompanying prominence (Howard 2015). In addition, Pevtsov et al. (2012) reported two CMEs on August 9, 2001 and March 3, 2011, which erupted from relatively empty filament channels without filament material. Finally, MCs are three-dimensional structures, which are sampled by a spacecraft from a single vantage point, so the chances of missing the filament material are high (Gopalswamy et al. 2006, Sharma \& Srivastava 2012; Manchester et al. 2017).

In summary, we examined 76 MCs observed by ACE to search for cold prominence materials using the criteria of unusual $O^{5+}$ and (or) $Fe^{6+}$ abundances, and found many more MC events that contained prominence material. The result indicates that the unusual $O^{5+}$ and (or) $Fe^{6+}$ abundances may be simple and reliable criteria to investigate prominence materials in the interplanetary medium.

\begin{acknowledgements}

The authors acknowledge support from the  NSFC under grant Nos. 41674170. This work is also supported in part by the Plan for Scientific Innovation Talent of Henan Province under grant Nos. 174100510019. The authors thank NASA/GSFC for the use of data from ACE; these data can be obtained freely from the Coordinated Data Analysis Web (http://cdaweb.gsfc.nasa.gov/cdaweb/istp$\_$public/). The authors also thank the staff from CHIANTI for allowing us to use their database. CHIANTI is a collaborative project involving George Mason University, the University of Michigan (USA), and the University of Cambridge (UK). All data and programs are also freely available at the CHIANTI Web site (http://www.chiantidatabase.org).
\end{acknowledgements}

\begin{table}
\caption{Cold material regions within magnetic clouds and the maximum values of $O^{5+}$ and $Fe^{6+}$ fractions within the cold material regions.}
\begin{flushleft}
\begin{tabular}{lccccccc}
\hline
\multicolumn{1}{c}{NO.}&Front boundary$^{a}$&Rear boundary$^{b}$&Start of cold material$^{c}$&End of cold material$^{d}$&$O^{5+~e}$&$Fe^{6+~f}$\\
\hline
001&    1998/02/04 02:16&       1998/02/05 22:27&       1998/02/04 12:16&       1998/02/04 16:18&  0.058&  0.119\\
002&    1998/02/04 02:16&       1998/02/05 22:27&       1998/02/05 03:01&       1998/02/05 22:27&  0.038&  0.339\\
003&    1998/02/15 00:48&       1998/02/15 21:48&       1998/02/15 09:02&       1998/02/15 17:07&  0.062&  0.133\\
004&    1998/03/04 14:30&       1998/03/06 05:34&       1998/03/05 20:04&       1998/03/06 02:30&  0.062&  0.266\\
005&    1998/05/02 04:38&       1998/05/04 02:28&       1998/05/02 08:58&       1998/05/03 15:39&  0.452&  0.880\\
006&    1998/09/25 06:16&       1998/09/26 16:28&       1998/09/25 06:16&       1998/09/25 15:43&  0.194&  0.143\\
007&    1998/09/25 06:16&       1998/09/26 16:28&       1998/09/26 05:52&       1998/09/26 09:43&  0.141&  0.195\\
008&    1998/11/13 02:15&       1998/11/16 01:38&       1998/11/15 05:52&       1998/11/15 21:34&  0.071&  0.050\\
009&    1999/04/16 18:45&       1999/04/17 19:48&       1999/04/16 22:19&       1999/04/17 09:33&  0.052&  0.068\\
010&    1999/04/21 04:18&       1999/04/23 01:37&       1999/04/21 23:57&       1999/04/22 03:58&  0.050&  0.082\\
011&    1999/04/26 08:58&       1999/04/27 06:43&       1999/04/26 08:58&       1999/04/26 10:28&  0.023&  0.275\\
012&    1999/04/26 08:58&       1999/04/27 06:43&       1999/04/27 04:33&       1999/04/27 06:34&  0.078&  0.000\\
013&    1999/07/06 19:16&       1999/07/07 16:55&       1999/07/06 19:16&       1999/07/07 05:58&  0.089&  0.050\\
014&    2000/02/21 05:21&       2000/02/22 12:13&       2000/02/21 15:20&       2000/02/21 21:13&  0.141&  0.333\\
015&    2000/03/28 03:10&       2000/03/29 19:28&       2000/03/28 20:25&       2000/03/28 22:27&  0.161&  0.000\\
016&    2000/07/15 19:52&       2000/07/16 23:22&       2000/07/16 02:06&       2000/07/16 06:09&  0.065&  0.051\\
017&    2000/07/28 13:03&       2000/07/29 10:12&       2000/07/28 15:43&       2000/07/28 23:45&  0.051&  0.001\\
018&    2000/11/30 21:34&       2000/12/02 14:05&       2000/12/01 19:19&       2000/12/01 23:19&  0.073&  0.000\\
019&    2001/03/19 19:39&       2001/03/21 23:42&       2001/03/21 17:47&       2001/03/21 23:42&  0.011&  0.105\\
020&    2001/10/22 00:13&       2001/10/22 18:44&       2001/10/22 00:13&       2001/10/22 12:47&  0.258&  0.079\\
021&    2002/05/23 21:36&       2002/05/25 17:48&       2002/05/24 07:34&       2002/05/24 21:35&  0.261&  0.253\\
022&    2002/09/08 23:38&       2002/09/10 20:56&       2002/09/10 00:03&       2002/09/10 12:06&  0.073&  0.105\\
023&    2003/10/26 18:32&       2003/10/28 08:49&       2003/10/27 16:38&       2003/10/28 08:49&  0.336&  0.565\\
024&    2004/04/03 23:14&       2004/04/05 18:18&       2004/04/03 23:14&       2004/04/04 11:04&  0.043&  0.138\\
025&    2004/05/01 14:56&       2004/05/02 23:51&       2004/05/01 14:56&       2004/05/02 05:12&  0.071&  0.015\\
026&    2005/01/08 11:40&       2005/01/09 17:49&       2005/01/09 05:24&       2005/01/09 17:49&  0.245&  0.286\\
027&    2005/05/20 03:04&       2005/05/21 15:40&       2005/05/20 05:44&       2005/05/20 11:04&  0.225&  0.098\\
028&    2007/01/14 07:26&       2007/01/15 07:48&       2007/01/14 07:26&       2007/01/14 11:52&  0.119&  0.030\\
029&    2007/09/09 22:45&       2007/09/12 10:31&       2007/09/11 23:24&       2007/09/12 09:24&  0.164&  0.297\\
\hline
\end{tabular}
\end{flushleft}
a {The front boundary of the magnetic cloud (UT).}

b {The rear boundary of the magnetic cloud (UT).}

c {The beginning of the cold material regions within the magnetic cloud (UT).}

d {The end of the cold material regions within the magnetic cloud (UT).}

e {The maximum values of $O^{5+}$ fractions within cold material regions.}

f {The maximum values of $Fe^{6+}$ fractions within cold material regions.}
\end{table}

\end{document}